\documentclass[aps,prb,twocolumn,a4paper]{revtex4-1}
\usepackage[dvipdfmx]{graphicx}

\usepackage{amsmath,amsthm,amssymb}
\usepackage{amsmath,bm}
\usepackage{color}
\usepackage{ifthen}

\usepackage{multirow,bigdelim}

\newcount \commentout
\newcommand{\1}{\mbox{1}\hspace{-0.25em}\mbox{l}}

\newcommand{\sgn}{\mathrm{sgn}}

\newlength{\figwidth}
\newlength{\figlarge}
\begin{document}
\title{
Discriminant indicators with generalized inversion symmetry
}

\author{
Tsuneya Yoshida${}^1$
}
\author{
Ryo Okugawa${}^2$
}
\author{
Yasuhiro Hatsugai${}^1$
}
\affiliation{${}^1$ Department of Physics, University of Tsukuba, Ibaraki 305-8571, Japan}
\affiliation{${}^2$ Graduate School of Information Sciences, Tohoku University, Sendai 980-8579, Japan}
\date{\today}
\begin{abstract}
We propose indicators of the discriminant for systems with generalized inversion symmetry which are computed from data only at high-symmetry points in the Brillouin zone.
Our approach captures the exceptional points and their symmetry-protected variants without ambiguity arising from the reference energy, which is advantage over the previously known indicators for non-Hermitian systems.
As demonstrations, we systematically analyze $3\times 3$-Hamiltonians where the proper choice of the reference energy is not obvious.
\end{abstract}
\maketitle


\section{
Introduction
}
Non-Hermitian topology is studied as one of current hot topics attracting broad attention~\cite{Hatano_PRL96,CMBender_PRL98,Hu_nH_PRB11,Esaki_nH_PRB11,Bergholtz_Review19,Ashida_nHReview_arXiv20}.
When the Hermiticity is violated, systems exhibit novel topological phenomena due to the point-gap topology for which the non-Hermiticity is essential~\cite{Alvarez_nHSkin_PRB18,SYao_nHSkin-1D_PRL18,SYao_nHSkin-2D_PRL18,KFlore_nHSkin_PRL18,Yokomizo_BBC_PRL19,Gong_class_PRX18,Kawabata_gapped_PRX19,Zhou_gapped_class_PRB19}.
A representative example is the non-Hermitian skin effect which results in extreme sensitivity of eigenstates and eigenvectors to the presence/absence of boundaries~\cite{SYao_nHSkin-1D_PRL18,Lee_Skin19,Borgnia_ptGapPRL2020,Zhang_BECskin19,Okuma_BECskin19}. 
Another typical example is the emergence of exceptional points~\cite{TKato_EP_book1966,Rotter_EP_JPA09,Berry_EP_CzeJPhys2004,Heiss_EP_JPA12,HShen2017_non-Hermi} on which energy bands touch for both of the real- and imaginary-parts. 
The topology of exceptional points is further enriched by symmetry~\cite{Budich_SPERs_PRB19,Okugawa_SPERs_PRB19,Yoshida_SPERs_PRB19,Zhou_SPERs_Optica19}; symmetry-protected exceptional rings (SPERs) and symmetry-protected exceptional surfaces (SPESs) emerge in two and three dimensions.
Among studies of non-Hermitian systems, searching platforms of the above topological phenomena is considered to be an important issue in terms of applications and has been addressed extensively. 
So far, exceptional points and skin effects have been reported for open quantum systems~\cite{Jin_qEP_PRA09,Jose_DissSuperEP_SciRep16,TELeePRL16_Half_quantized,YXuPRL17_exceptional_ring,Lei_qEP_PRA19}, photonic systems~\cite{Guo_nHExp_PRL09,Ruter_nHExp_NatPhys10,Regensburger_nHExp_Nat12,Zhen_AcciEP_Nat15,Hassan_EP_PRL17,Takata_pSSH_PRL18,Zhou_FermiArcPH_Science18,Ozawa_TopoPhoto_RMP19,Xiao_nHSkin_Exp_NatPhys19}, mechanical systems~\cite{Yoshida_SPERs_mech19,Ghatak_Mech_nHskin_PNAS20,Scheibner_nHmech_PRL20}, electric circuits~\cite{Hofmann_ExpRecipSkin_19,Helbig_ExpSkin_19,Yoshida_MSkinPRR20}, quasi-particle spectrum~\cite{VKozii_nH_arXiv17,Zyuzin_nHEP_PRB18,Yoshida_EP_DMFT_PRB18,HShen2018quantum_osci,Papaji_nHEP_PRB19,Matsushita_ER_PRB19,Michishita_EP_PRL20,Yoshida_nHReview_PTEP20} and so on.

For further search of the platforms, indicators are considered to be powerful tools; in Hermitian case, symmetry-indicators are employed as efficient filters~\cite{Herring_Dirac_PR1937,Po_Indicator_NatComm2017,Ono_SymmInd-Hermichiral_PRB18,Po_indicatorRev_JOPCM20} of trivial topology and allow systematic search of topological insulators and semimetals. 
Symmetry-indicators are also extended to non-Hermitian cases~\cite{Okugawa_SymmInd-nH_PRB21,Vecsei_SymmInd-nH_PRB21,Shiozaki_SymmInd_PRB21,Okugawa_HOSkin_PRB20,Yoshida_nHFQHJ_PRR20,Tsubota_CorrInv_arXiv21}.
In particular, Refs.~\onlinecite{Okugawa_SymmInd-nH_PRB21,Vecsei_SymmInd-nH_PRB21,Shiozaki_SymmInd_PRB21} introduced indicators for exceptional points and the skin effects by analyzing the doubled Hermitian Hamiltonian which is composed of the non-Hermitian Hamiltonian and the reference energy.
However, these previously introduced indicators possess ambiguity arising from the reference energy, which prevent us from systematically searching the non-Hermitian topological phenomena.

In this paper, we propose another type of the indicators by focusing on the discriminant which allows to characterize exceptional points without input of the reference energy~\cite{Wojcik_DiscEP_PRB20,Yang_DiscEP_PRL21,Delplace_Resul_arXiv21}.
Our discriminant indicators allow systematic search of exceptional points and symmetry-protected variants (i.e., SPERs and SPESs) for systems with generalized inversion symmetry.
The above advantage is demonstrated by analyzing non-Hermitian Hamiltonians where a proper choice of the reference energy is not obvious.

The rest of this paper is organized as follows. After a brief review of the discriminant in Sec.~\ref{sec: rev of disc}, we introduce the discriminant indicator for systems with generalized inversion symmetry in Sec.~\ref{sec: disc num for EP} where a numerical simulation is also performed.
In Sec.~\ref{sec: disc inv for SPERs}, we propose the discriminant indicator for SPERs and SPESs for generalized time-reversal and inversion symmetric systems.
A short summary is given in Sec.~\ref{summary}.
Appendices are devoted to details of the discriminant (Appendix~\ref{sec: Disc app}), relations between symmetry indicators and discriminant indicators (Appendix~\ref{sec: symm ind app}), and details of symmetry constraints (Appendix~\ref{sec: symm const app}).

\section{
Brief review of the discriminant and exceptional points
}
\label{sec: rev of disc}

Consider a two-dimensional system which is described by a $2\times 2$-Hamiltonian $H(\bm{k})$.
Exceptional points in this system is characterized by the following winding number~\cite{HShen2017_non-Hermi}
\begin{eqnarray}
\nu_{2\times2} &=& \oint \frac{d \bm{k}}{2\pi i} \cdot \nabla_{\bm{k}} \log[\epsilon_1(\bm{k})-\epsilon_2(\bm{k})],
\end{eqnarray}
where $\epsilon_n(\bm{k})$ ($n=1,2$) denote eigenvalues of the Hamiltonian.
The integral is taken along a closed path.
The operator $\nabla_{\bm{k}}$ is defined as $\nabla_{\bm{k}}:=(\partial_{k_x},\partial_{k_y})$. 
Here, $\partial_{k_x}$ ($\partial_{k_y}$) denotes derivative respect to $k_x$ ($k_y$) where $k_x$ and $k_y$ are $x$- and $y$- components of the momentum respectively.

The discriminant number is introduced as an generalized version of $\nu_{2\times2}$ to an $N\times N$-Hamiltonian $H(\bm{k})$~\cite{Wojcik_DiscEP_PRB20,Yang_DiscEP_PRL21,Delplace_Resul_arXiv21}.
\begin{eqnarray}
\label{eq: defs of disc number}
\nu &=& \oint \frac{d \bm{k}}{2\pi i} \cdot \nabla_{\bm{k}} \log \Delta(\bm{k}).
\end{eqnarray}
Here, $\Delta(\bm{k})$ denotes the discriminant of the polynomial $\det[H(\bm{k})-E\1]=a_N (-E)^N+a_{N-1}(-E)^{N-1}+\cdots+a_0$ ($a_l\in \mathbb{C}$) which is defined as
\begin{eqnarray}
\label{eq: Defs of disc}
\Delta(\bm{k})&:=& (-1)^{N(N-1)/2}\prod_{n\neq n'}(\epsilon_n-\epsilon_{n'}),
\end{eqnarray}
with $\epsilon_n$ being roots of the polynomial [i.e., eigenvalues of $H(\bm{k})$].
We note that the discriminant is written in terms of $a_l$'s (for more details see Appendix~\ref{sec: Disc app}).
If the discriminant number takes an non-trivial value for a closed path in the Brillouin zone (BZ), $\Delta(\bm{k})$ vanishes at a point inside of the loop.
Therefore, at least two bands touch at this point [see Eq.~(\ref{eq: Defs of disc})].

Another generalization is the following winding number~\cite{Kawabata_gapless_PRL19}
\begin{eqnarray}
W(E_{\mathrm{ref}}) &=& \oint \frac{d \bm{k}}{2\pi i} \cdot \nabla_{\bm{k}} \log \det[H(\bm{k})-E_{\mathrm{ref}}\1],
\end{eqnarray}
where $E_{\mathrm{ref}}$ is chosen as the energy at which an exceptional point emerge.
In other words, for the characterization by $W(E_{\mathrm{ref}})$, one needs to know a proper choice of $E_{\mathrm{ref}}$ by diagonalizing the Hamiltonian.
The advantage of the discriminant number is that one does not need to know energy at which two bands touch apriori.
\section{
Indicators for generalized inversion symmetric systems
}
\label{sec: disc num for EP}
Consider a system with generalized inversion symmetry which is written as
\begin{eqnarray}
\label{eq: H inv dag}
U_{\mathrm{I}} H(\bm{k}) U^\dagger_{\mathrm{I}}&=& H^\dagger(-\bm{k}),
\end{eqnarray}
where $U_{\mathrm{I}}$ is a Hermitian and unitary matrix (i.e., $U_{\mathrm{I}}U^\dagger_{\mathrm{I}}=\1$ and $U_{\mathrm{I}}^{\dagger}=U_{\mathrm{I}}$).

Equation~(\ref{eq: H inv dag}) results in the following constraint on the discriminant:
\begin{eqnarray}
\label{eq: Delta inv dag}
\Delta(\bm{k})&=& \Delta^*(-\bm{k}),
\end{eqnarray}
which is proven in Sec.~\ref{sec: EP inv 2D} [see Eq.~(\ref{eq: defs LR eigen})-Eq.~(\ref{eq: e^*(k)=e(-k) inv})].

As we see below, the symmetry constraint on the discriminant allows to define the indicator~(\ref{eq: nu const inv 2D}) [Eq.~(\ref{eq: nu const inv 3D})] which captures exceptional points [exceptional loops] in two dimensions [three dimensions].

\subsection{
Two dimensions
}
\label{sec: EP inv 2D}
\begin{figure}[!h]
\begin{minipage}{1\hsize}
\begin{center}
\includegraphics[width=1\hsize,clip]{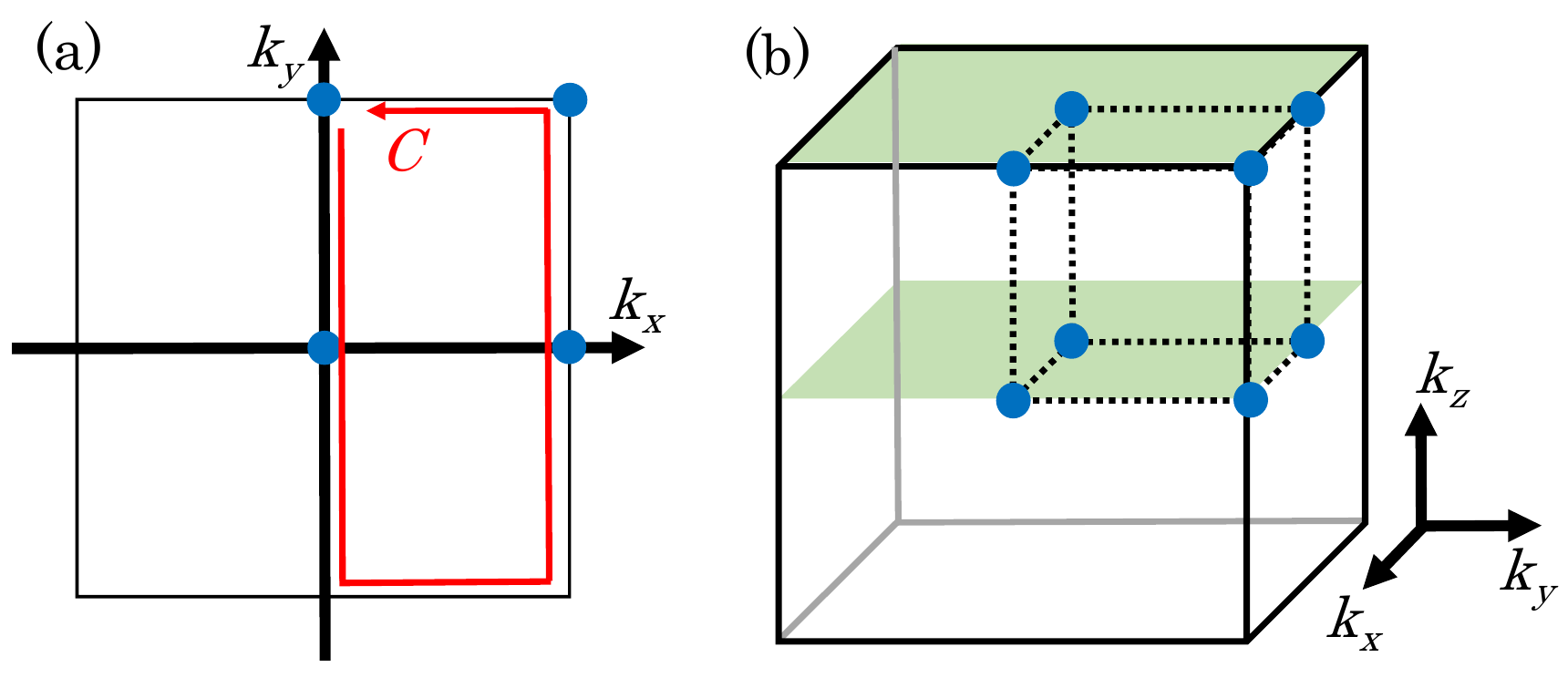}
\end{center}
\end{minipage}
\caption{
(a) [(b)]: Sketch of the BZ of the square [cubic] lattice, $-\pi < k_{\mu}\leq \pi$ with $\mu=x,y$ [$\mu=x,y,z$]. 
Blue dots denotes TRIM which are invariant under the transformation $\bm{k} \to -\bm{k}$. Red lines in panel (a) denotes a closed path where the integral in Eq.~(\ref{eq: defs of disc number}) is computed.
}
\label{fig: BZ}
\end{figure}

Consider the discriminant number $\nu$ computed along the closed path illustrated in Fig.~\ref{fig: BZ}(a).
The generalized inversion symmetry imposes the following constraint on the parity of $\nu$
\begin{eqnarray}
\label{eq: nu const inv 2D}
(-1)^\nu &=& \zeta_{\mathrm{I,2D}}:=\prod_{\bm{\Gamma}_j\in \mathrm{TRIM}} \sgn\Delta(\bm{\Gamma}_j),
\end{eqnarray}
where $\mathrm{TRIM}$ denotes time-reversal symmetric momenta
\begin{eqnarray}
\mathrm{TRIM}&=& \{(0,0),\, (\pi,0), \, (0,\pi),\, (\pi,\pi) \}.
\end{eqnarray}
We note that $\Delta(\bm{\Gamma}_j)$ ($\bm{\Gamma}_j\in \mathrm{TRIM}$) are real because of Eq.~(\ref{eq: Delta inv dag}).
Equation~(\ref{eq: nu const inv 2D}) means that $\zeta_{\mathrm{I},2D}$ taking $-1$ indicates the emergence of exceptional points.

Eigenvalues of the inversion operator do not explicitly appear in Eq.~(\ref{eq: nu const inv 2D}), while the indicator computed from the eigenvalues of symmetry operators in Hermitian case.
We note, however, that the product of parity eigenvalues is computed in right hand side of Eq.~(\ref{eq: nu const inv 2D}), which is discussed in Appendix~\ref{sec: symm ind app}.

Equation~(\ref{eq: nu const inv 2D}) can be proven by taking into account the symmetry constraint [Eq.~(\ref{eq: H inv dag})].
As a first step, we prove Eq.~(\ref{eq: Delta inv dag}).
Let $|R_n(\bm{k})\rangle$ and $|L_n(\bm{k})\rangle$ be right and left eigenvectors of $H(\bm{k})$
\begin{subequations}
\label{eq: defs LR eigen}
\begin{eqnarray}
H(\bm{k}) |R_n(\bm{k})\rangle &=&  \epsilon_n(\bm{k}) |R_n(\bm{k})\rangle, \\
H^\dagger(\bm{k}) |L_n(\bm{k})\rangle &=&  \epsilon^*_n(\bm{k}) |L_n(\bm{k})\rangle.
\end{eqnarray}
\end{subequations}
Then, we have
\begin{eqnarray}
\label{eq: HI|L>}
H(-\bm{k}) U_{\mathrm{I}} |L_n(\bm{k})\rangle&=& U_{\mathrm{I}} H^\dagger(\bm{k}) |L_n(\bm{k})\rangle \nonumber \\
                               &=& \epsilon^*_{n}(\bm{k}) U_{\mathrm{I}}  |L_n(\bm{k})\rangle,
\end{eqnarray}
which results in
\begin{eqnarray}
\label{eq: e^*(k)=e(-k) inv}
\epsilon^*_n(-\bm{k})&=&  \epsilon_n(\bm{k}).
\end{eqnarray}
Combining Eq.~(\ref{eq: Defs of disc}) and Eq.~(\ref{eq: e^*(k)=e(-k) inv}), we obtain Eq.~(\ref{eq: Delta inv dag}).

Now, we evaluate the discriminant number along the path illustrated in Fig.~\ref{fig: BZ}(a).
Firstly, we decompose the integral into two parts
\begin{eqnarray}
\label{eq: nu decomp 2D}
\nu &=& \nu_0-\nu_\pi,
\end{eqnarray}
with 
\begin{eqnarray}
2\pi i \nu_{k_0} &=& \int^{\pi}_{-\pi}\! dk_y \partial_{k_y} \log\Delta(k_0,k_y),
\end{eqnarray}
$\Delta(k_x,k_y):=\Delta(\bm{k})$, and $k_0$ taking $0$ or $\pi$.
Because the inversion is closed for one-dimensional subsystems specified by $k_{0}=0,\pi$, the integral can be simplified as follows~\cite{Hughes_Inv_PRB11,Hatsugai_ZQBerry_EPL11}:
\begin{eqnarray}
&& 2i\pi \nu_{k_{0}} \nonumber \\
&&= \int^{\pi}_{0}\! dk_y \partial_{k_y} \log \Delta(k_0,k_y) + \int^{0}_{-\pi}\! dk_y \partial_{k_y} \log \Delta(k_0,k_y) \nonumber \\
&&= \int^{\pi}_{0}\! dk_y \partial_{k_y} \log \Delta(k_0,k_y) + \int^{0}_{-\pi}\! dk_y \partial_{k_y} \log \Delta^*(k_0,-k_y) \nonumber \\
&&= \int^{\pi}_{0}\! dk_y \partial_{k_y} \log \Delta(k_0,k_y) - \int^{\pi}_{0}\! dp_y \partial_{p_y} \log \Delta^*(k_0,p_y) \nonumber \\
&&= 2 i \mathrm{Im} \int^{\pi}_{0}\! dk_y \partial_{k_y} \log \Delta(k_0,k_y) \nonumber\\
&&= 2 i \left[ \mathrm{Arg}\Delta(k_{0},\pi)-\mathrm{Arg}\Delta(k_{0},0) +2\pi N_0\right],
\end{eqnarray}
with an integer $N_0$.
Here, from the second to the third line, we have used Eq.~(\ref{eq: Delta inv dag}).
Taking a principal value [$-\pi<\mathrm{Arg}\Delta(\bm{k})\leq \pi$] results in $N_0$ ambiguity in the last line.
The above results indicate the following relation 
\begin{eqnarray}
\label{eq: nu const inv 1D}
e^{i\pi \nu_{k_{0}} }&=& e^{i \left[ \mathrm{Arg}\Delta(k_{0},\pi)-\mathrm{Arg}\Delta(k_{0},0)\right]} \nonumber \\
                     &=& \mathrm{sgn}\Delta(k_{0},\pi)\mathrm{sgn}\Delta(k_{0},0).
\end{eqnarray}
Putting Eqs.~(\ref{eq: nu decomp 2D})~and~(\ref{eq: nu const inv 1D}) together, we end up with Eq.~(\ref{eq: nu const inv 2D}).

In the above, we have assumed that $\Delta(\bm{k})$ is non-zero along the path denoted by red lines in Fig.~\ref{fig: BZ}(a). We note, however, that zeros of $\Delta(\bm{k})$ on the path but away from $\mathrm{TRIM}$ does not change the conclusion. 
This is because we can take another path where the inversion is closed and $\Delta(\bm{k})$ is finite~\cite{Kim_Berryinv_PRL15}.

\subsection{
Three dimensions
}

Generically, exceptional points form a loop in the three-dimensional BZ [$-\pi < k_{\mu}\leq \pi $ ($\mu=x,y,z$)]~\cite{YXuPRL17_exceptional_ring}.
In this case, focuing on $k_z=0,\pi$, the problem is reduced to the case discussed in Sec.~\ref{sec: EP inv 2D} [see green planes in Fig.~\ref{fig: BZ}(b)].
Thus, we can use the following indicator to detect the exceptional loop
\begin{eqnarray}
\label{eq: nu const inv 3D}
\zeta_{\mathrm{I,3D}} &:=&\prod_{\bm{\Gamma}_j\in \mathrm{TRIM}}\mathrm{sgn}\Delta(\bm{\Gamma}_j),
\end{eqnarray}
where $\mathrm{TRIM}$ denotes time-reversal symmetric momenta~\cite{EP3Dind_ftnt}
\begin{eqnarray}
\mathrm{TRIM}&=& \{(0,0,0),\, (\pi,0,0), \, (0,\pi,0),\, (\pi,\pi,0), \nonumber \\
              &&   (0,0,\pi),\, (\pi,0,\pi), \, (0,\pi,\pi),\, (\pi,\pi,\pi) \}.
\end{eqnarray}

\subsection{
Application to a toy model
}
Now, we apply the above approach to a toy model in two dimensions where a proper value of the reference energy is not obvious.
The Hamiltonian reads
\begin{eqnarray}
H(\bm{k}) 
&=& 
\xi(\bm{k})
\left(
\begin{array}{ccc}
\frac{1}{2} & 0  &  0  \\
     0      & 0  &  1\\
     0      & 1  &  0
\end{array}
\right)
+(i\gamma+\sin k_x)
\left(
\begin{array}{ccc}
0 & 0 & 0  \\
0 & 0 &-i  \\
0 & i & 0
\end{array}
\right)\nonumber \\
&&+(i\gamma+\sin k_y)
\left(
\begin{array}{ccc}
0 &  0  & 0 \\
0 &  1  & 0  \\
0 &  0  & -1
\end{array}
\right) 
+\xi'(\bm{k})
\left(
\begin{array}{ccc}
0             & \frac{1}{2} & -\frac{1}{2}  \\
\frac{1}{2}   & 0 & i  \\
-\frac{1}{2}  & i & 0
\end{array}
\right) \nonumber \\
&&
+\xi''(\bm{k})
\left(
\begin{array}{ccc}
1 & 0 & 0  \\
0 & 0 & 0 \\
0 & 0 & 0
\end{array}
\right),
\end{eqnarray}
with $\xi(\bm{k})=-2t(\cos k_x+\cos k_y)-m$, $\xi'(\bm{k})=0.3i\cos(k_x+k_y)+0.5\sin(k_x-k_y)$, and $\xi''(\bm{k})=0.3i\sin(k_x-k_y)+2\cos(k_x+k_y)$. 
Parameters $m$, $t$, and $\gamma$ are real numbers.
This Hamiltonian preserves the generalized inversion symmetry [see Eq.~(\ref{eq: H inv dag})] with
\begin{eqnarray}
\label{eq: 3x3 inv}
U_{\mathrm{I}}&=& 
\left(
\begin{array}{ccc}
1  & 0  & 0 \\
0  & 0  & 1 \\
0  & 1  & 0
\end{array}
\right).
\end{eqnarray}
We note that the above Hamiltonian with generalized inversion symmetry can be obtained by
\begin{eqnarray}
H(\bm{k})&=& \frac{1}{2} \left[ H_0(\bm{k})+U_{\mathrm{I}}H^\dagger_0(-\bm{k})U^{\dagger}_{\mathrm{I}} \right],
\end{eqnarray}
with a general Hamiltonian $H_0(\bm{k})$. 

\begin{figure}[!h]
\begin{minipage}{1\hsize}
\begin{center}
\includegraphics[width=1\hsize,clip]{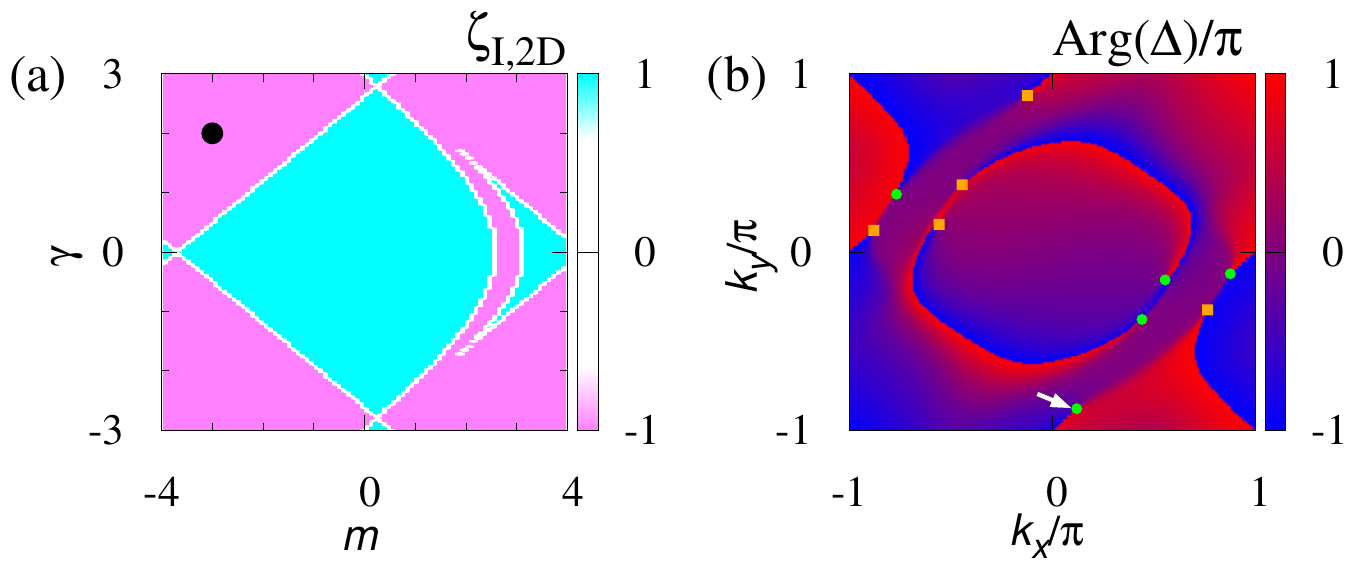}
\end{center}
\end{minipage}
\begin{minipage}{1\hsize}
\begin{center}
\includegraphics[width=1\hsize,clip]{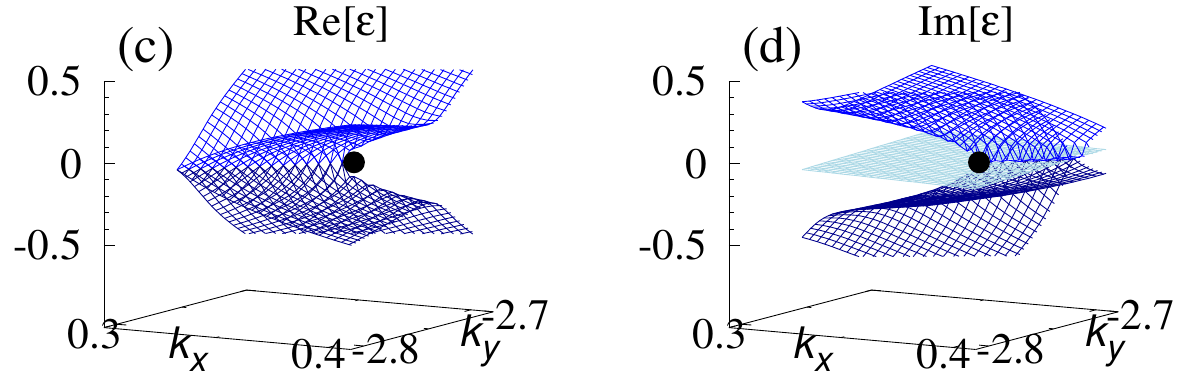}
\end{center}
\end{minipage}
\caption{
(a): Discriminant indicator [$\zeta_{\mathrm{I,2D}}=(-1)^{\nu}$] for $t=0.5$.
(b): Argument of $\Delta(\bm{k})$ normalized by $\pi$ in the BZ. 
The orange (green) symbols denote exceptional points with $\nu=1$ ($-1$).
(c) [(d)]: The real- (imaginary-) part of the energy bands around $\bm{k}_{\mathrm{EP}1}$.
These data are obtained for $(m,t,\gamma)=(-3,0.5,2)$ [see the black dot in panel (a)].
}
\label{fig: arg 2D EP}
\end{figure}

Figure~\ref{fig: arg 2D EP}(a) plots a phase diagram against $m$ and $\gamma$ for $t=0.5$. 
It indicates that exceptional points emerge in the region colored with pink.
For the sake of concreteness, let us focus on the case of $(m,t,\gamma)=(-3,0.5,2)$ [see the black dot in Fig.~\ref{fig: arg 2D EP}(a)].
Figure~\ref{fig: arg 2D EP}(b) is a color plot of $\mathrm{Arg}\Delta(\bm{k})/\pi$ in the momentum space for $(m,t,\gamma)=(-3,0.5,2)$.
It indicates that the discriminant number computed along the path illustrated in Fig.~\ref{fig: BZ}(a) takes $-3$, which is consistent with the discriminant indicator taking $-1$.
Exceptional points emerge at points denoted by orange and green dots in Fig.~\ref{fig: arg 2D EP}(b), which can be confirmed in Figs.~\ref{fig: arg 2D EP}(c)~and~\ref{fig: arg 2D EP}(d).
The above results indicate that the discriminant indicator successfully captures the exceptional points.

We finish this part with a remark on the indicator in previous works~~\cite{Okugawa_SymmInd-nH_PRB21,Vecsei_SymmInd-nH_PRB21,Shiozaki_SymmInd_PRB21},
\begin{eqnarray}
\label{eq: ind with Eref for EP}
(-1)^{W(E_{\mathrm{ref}})}&=& \prod_{\bm{\Gamma}_j\in \mathrm{TRIM}} \mathrm{sgn} \left( \mathrm{det}[H(\bm{\Gamma}_j)-E_{\mathrm{ref}}\1] \right).
\end{eqnarray}
To compute this indicator, one needs to know a proper choice of $E_{\mathrm{ref}}$ by analyzing the band structure.
In addition, changing a parameter (e.g., $m$ or $\gamma$) shifts the energy where two bands touch.
These behaviors prevent us from systematically searching the exceptional points based on Eq.~(\ref{eq: ind with Eref for EP}).
In contrast, one can carry out a systematic analysis by discriminant indicators [Eqs.~(\ref{eq: nu const inv 2D})~and ~(\ref{eq: nu const inv 3D})].

\section{
Indicators for generalized time-reversal and inversion symmetric systems
}
\label{sec: disc inv for SPERs}

\subsection{
Two and three dimensions
}

Consider generalized time-reversal and inversion symmetric systems
\begin{subequations}
\label{eq: THT=H^T and IHI=Hdag}
\begin{eqnarray}
\label{eq: THT=H^T}
U_{\mathrm{T}} H^T(\bm{k}) U^{\dagger}_{\mathrm{T}}&=& H(-\bm{k}), \\
\label{eq: IHI=Hdag}
U_{\mathrm{I}} H^\dagger (\bm{k}) U^{\dagger}_{\mathrm{I}}&=& H(-\bm{k}),
\end{eqnarray}
\end{subequations}
with unitary operators $U_{\mathrm{T}}$ and $U_{\mathrm{I}}$ satisfying $U_{\mathrm{T}}U^*_{\mathrm{T}}=\1$ and $U_{\mathrm{I}}U_{\mathrm{I}}=\1$, respectively.
In Ref.~\onlinecite{Kawabata_gapped_PRX19}, symmetry described by Eq.~(22a) is denoted as $\mathrm{TRS}^\dagger$.
In two (thee) dimensions, the systems host SPERs (SPESs) which are characterized by the sign of the discriminant~\cite{Delplace_Resul_arXiv21,Yoshida_nHgame_arXiv21}.

In order to introduce an indicator for these symmetry-protected topological properties, 
let us start with the following constraints imposed by symmetry Eq.~(\ref{eq: THT=H^T and IHI=Hdag}):
\begin{subequations}
\label{eq: const. on Delta SPER}
\begin{eqnarray}
\Delta(\bm{k})&=& \Delta^*(\bm{k}), \\
\Delta(\bm{k})&=& \Delta(-\bm{k}), 
\end{eqnarray}
\end{subequations}
which are proven in Sec.~\ref{sec: proof of const on Delta SPER}.
Because the real function $\Delta(\bm{k})$ is continuous and inversion symmetric, we can introduce the following indicator for SPERs and SPESs,
\begin{eqnarray}
\label{eq: ind for SPERs}
\zeta_{\mathrm{TI}}&=& \prod_{\bm{\Gamma}_j\in \mathrm{TRIM}} \mathrm{sgn}\Delta(\bm{\Gamma}_j).
\end{eqnarray}
When $\zeta_{\mathrm{TI}}$ takes $-1$, SPERs (SPESs) emerge in the two- (three-) dimensional BZ.

The above indicator also works for systems with  another type of time-reversal and inversion symmetry
\begin{subequations}
\label{eq: THT=H^* and IHI=H}
\begin{eqnarray}
\label{eq: THT=H^*}
U_{\mathrm{T}} H^*(\bm{k}) U^{\dagger}_{\mathrm{T}}&=& H(-\bm{k}), \\
\label{eq: IHI=H}
U_{\mathrm{I}} H (\bm{k}) U^{\dagger}_{\mathrm{I}}&=& H(-\bm{k}),
\end{eqnarray}
\end{subequations}
with $U_{\mathrm{T}}U^*_{\mathrm{T}}=\1$ because the above constraint results in Eq.~(\ref{eq: const. on Delta SPER}) (see Appendix~\ref{sec: symm const app}).

\subsection{
Proof of Eq.~(\ref{eq: const. on Delta SPER})
}
\label{sec: proof of const on Delta SPER}

When the Hamiltonian satisfies Eq.~(\ref{eq: THT=H^T and IHI=Hdag}), we obtain Eq.~(\ref{eq: const. on Delta SPER}).
Firstly, we note that energy eigenvalues satisfy
\begin{eqnarray}
\label{eq: e(-k) = e(k)}
\epsilon_n(-\bm{k})&=& \epsilon_n(\bm{k}),
\end{eqnarray}
in the presence of the generalized time-reversal symmetry, which is proven as follows.
Let $|L_n(\bm{k})\rangle$ be left eigenvectors of $H(\bm{k})$ with eigenvalues $\epsilon_n(\bm{k})$ ($n=1,2,3,\ldots$) [see Eq.~(\ref{eq: defs LR eigen}b)].
Then, we have $U_{\mathrm{T}}\mathcal{K}|L_n(\bm{k})\rangle$ as right eigenvectors of $H(-\bm{k})$ with eigenvalues $\epsilon_n(\bm{k})$
\begin{eqnarray}
H(-\bm{k})U_{\mathrm{T}}\mathcal{K}|L_n(\bm{k})\rangle &=& U_{\mathrm{T}}H^T(\bm{k})\mathcal{K}|L_n(\bm{k})\rangle \nonumber \\
                                          &=& U_{\mathrm{T}}\mathcal{K} H^\dagger(\bm{k})|L_n(\bm{k})\rangle \nonumber \\
                                          &=& \epsilon_n(\bm{k}) U_{\mathrm{T}}\mathcal{K}  |L_n(\bm{k})\rangle,
\end{eqnarray}
where $\mathcal{K}$ is a complex conjugation operator.
This relation prove that Eq.~(\ref{eq: e(-k) = e(k)}) holds.
Equation~(\ref{eq: e(-k) = e(k)}) results in Eq.~(\ref{eq: const. on Delta SPER}b), which can be seen by noting that $\Delta(\bm{k})$ is computed from the energy eigenvalues $\epsilon_n(\bm{k})$ [see Eq.~(\ref{eq: Defs of disc})].
As we have seen in Sec.~\ref{sec: EP inv 2D}, the generalized inversion symmetry results in $\Delta(\bm{k})=\Delta^*(-\bm{k})$ [see Eq.~(\ref{eq: Delta inv dag})].
Thus, combining this equation and Eq.~(\ref{eq: const. on Delta SPER}b), we obtain Eq.~(\ref{eq: const. on Delta SPER}a).

\subsection{
Application to a toy model
}
We demonstrate that the indicator [Eq.~(\ref{eq: ind for SPERs})] captures the SPERs in two dimensions even when energy where two band touch depends on momentum.
Let us analyze the following $3\times 3$-Hamiltonian $H(\bm{k})$ in two dimensions
\begin{eqnarray}
H(\bm{k})&=& 
i\xi(\bm{k})
\left(
\begin{array}{ccc}
     0      & 0  &  0  \\
     0      & 1  &  0\\
     0      & 0  &  -1
\end{array}
\right)
+it\sin k_x 
\left(
\begin{array}{ccc}
     0      & 0   &  0  \\
     0      & 0   &  1  \\
     0      & -1  &  0
\end{array}
\right)
\nonumber \\
&&+t\sin k_y 
\left(
\begin{array}{ccc}
     0      & 1   &  1  \\
    -1      & 0   &  0  \\
    -1      & 0   &  0
\end{array}
\right) 
+\gamma \cos k_x 
\left(
\begin{array}{ccc}
     1      & 0   &  0  \\
     0      & 0   &  0  \\
     0      & 0   &  0
\end{array}
\right)
\nonumber \\
&&+\gamma \cos (k_x+k_y) 
\left(
\begin{array}{ccc}
     0      & 1   &  1  \\
     1      & 0   &  0  \\
     1      & 0   &  0
\end{array}
\right),
\end{eqnarray}
with $\xi(\bm{k})=\cos k_x+\cos k_y-m$.
This Hamiltonian satisfies Eqs.~(\ref{eq: THT=H^T})~and~(\ref{eq: IHI=Hdag}) with $U_{\mathrm{T}}=\1$ and $U_{\mathrm{I}}$ defined in Eq.~(\ref{eq: 3x3 inv}).

\begin{figure}[!h]
\begin{minipage}{0.48\hsize}
\begin{center}
\includegraphics[width=1\hsize,clip]{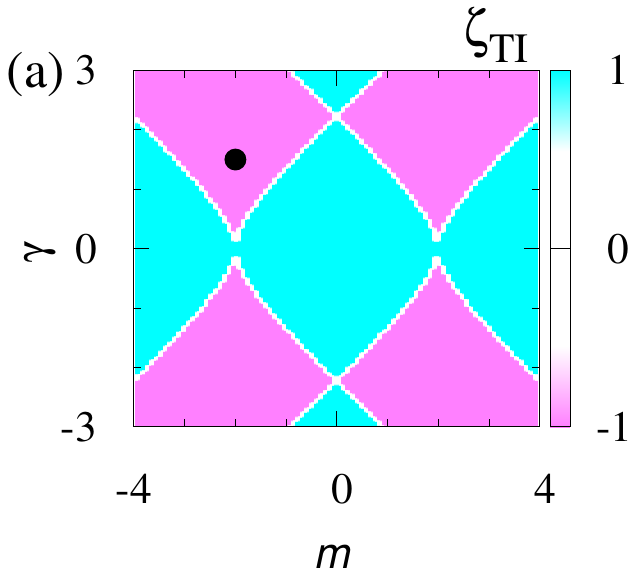}
\end{center}
\end{minipage}
\begin{minipage}{0.48\hsize}
\begin{center}
\includegraphics[width=1\hsize,clip]{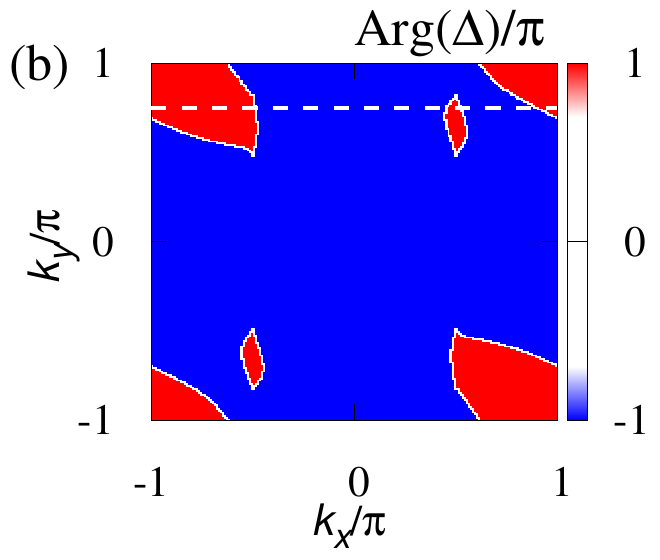}
\end{center}
\end{minipage}
\begin{minipage}{1\hsize}
\begin{center}
\includegraphics[width=1\hsize,clip]{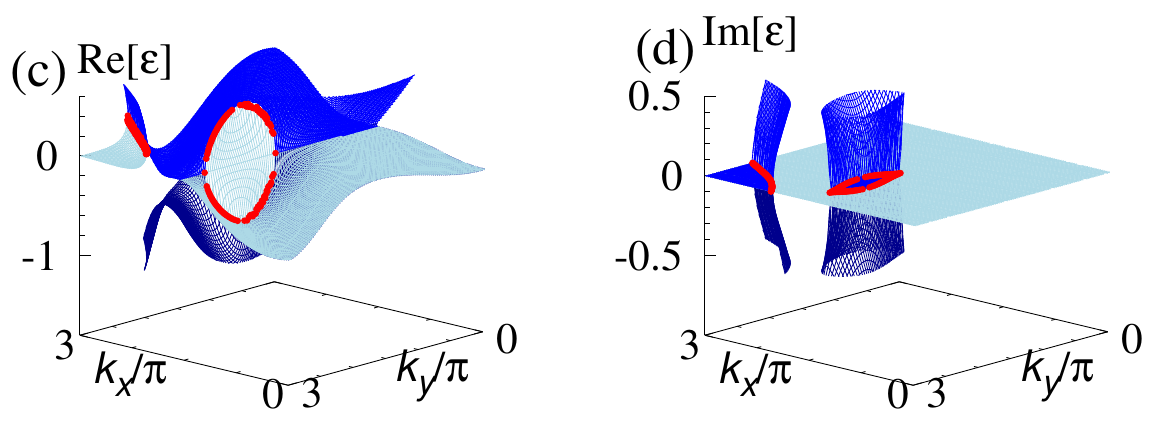}
\end{center}
\end{minipage}
\begin{minipage}{1\hsize}
\begin{center}
\includegraphics[width=1\hsize,clip]{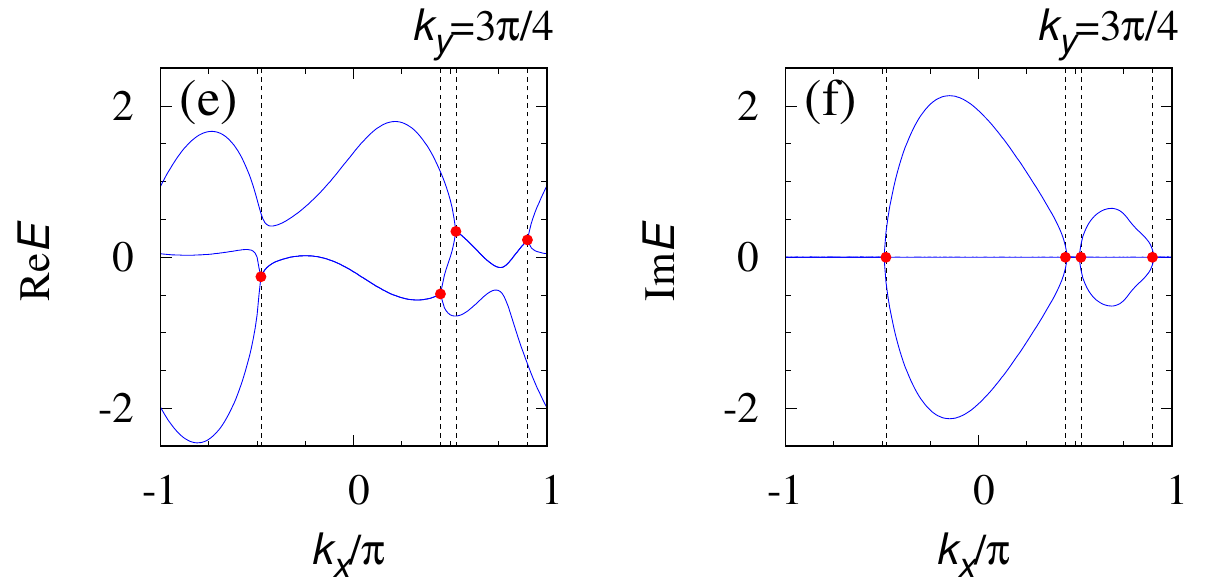}
\end{center}
\end{minipage}
\caption{
(a): The indicator for $t=0.5$. The black dot denotes a parameter set $(m,\gamma)=(-2,1.5)$.
(b): Sign of the discriminant $\Delta(\bm{k})$ in the BZ.
(c) [(d)]: The real- (imaginary-) part of the energy bands for $0 \leq  k_x \leq \pi$ and $0 \leq k_y \leq \pi$.
(e) [(f)]: The real- (imaginary-) part of the energy bands for $-\pi \leq k_x \leq \pi$ and $k_y=3\pi/4$ [see also dashed line in panel (b)].
Panels (b)-(f) are obtained for $(m,t,\gamma)=(-2,0.5,1.5)$.
}
\label{fig: arg 2D SPER}
\end{figure}
The phase diagram is shown in Fig.~\ref{fig: arg 2D SPER}(a). 
In the region colored with pink, the indicator takes $-1$ and predicts the emergence of SPERs.
Figure~\ref{fig: arg 2D SPER}(b) displays $\mathrm{sgn}\Delta(\bm{k})$ for $(m,t,\gamma)=(-2,0.5,1.5)$ [see the black dot in Fig.~\ref{fig: arg 2D SPER}(a)].
In this figure, we can see regions where $\mathrm{sgn}\Delta(\bm{k})$ takes $1$ or $-1$. On the boundaries, the discriminant $\Delta(\bm{k})$ becomes zero.
Correspondingly, the two bands touch as shown in Fig.~\ref{fig: arg 2D SPER}(c)~and~\ref{fig: arg 2D SPER}(d), which indicates the emergence of SPERs.

We finish this part with a remark on an indicator 
\begin{eqnarray}
z_{\mathrm{TI}} &=&
\prod_{\bm{\Gamma}_j \in \mathrm{TRIM}} \mathrm{sgn}\left( \mathrm{det}\left[ H(\bm{\Gamma}_j)-E_{\mathrm{ref}}\1 \right]  \right),
\end{eqnarray}
with $E_{\mathrm{ref}} \in \mathbb{R}$.
Detecting the above SPERs with this indicator is difficult because the energy where two bounds touch depends on the momentum [see Figs.~\ref{fig: arg 2D SPER}(e)~and~\ref{fig: arg 2D SPER}(f)].
Figure~\ref{fig: arg 2D SPER} indicates that the indicator defined in Eq.~(\ref{eq: ind for SPERs}) captures SPERs even in this case.

\section{
Summary
}
\label{summary}
In this paper, we have proposed the discriminant indicator for systems with generalized inversion symmetry. 
In contrast to the previously introduced indicators, our approach captures exceptional points without ambiguity arising from the choice of the reference energy. 
A similar indicator is also introduced for SPERs (SPERs) in two- (three-) dimensional systems with generalized inversion and time-reversal symmetry.
Applying the discriminant indicators to $3\times3$-Hamiltonians, we have demonstrated that our approach successfully captures exceptional points and SPERs even when a proper choice of the reference energy is not obvious.

We finish this paper with two remarks. 
Firstly, we note that the previously introduced indicators~\cite{Okugawa_SymmInd-nH_PRB21,Vecsei_SymmInd-nH_PRB21,Shiozaki_SymmInd_PRB21} are also applicable to skin effects, while our indicators focus only on exceptional points and their symmetry-protected variants. 
Secondly, we note that while indicators for the line-gap topology has been discussed for correlated systems~\cite{Yoshida_nHFQHJ_PRR20,Tsubota_CorrInv_arXiv21}, indicators for the point-gap topology have not been discussed so far. 
Introducing indicators for correlated systems is expected to accelerate search of the non-Hermitian topology of such systems which is left as a future work to be addressed.
%
%
\section*{
Acknowledgments
}
T. Y. and Y. H. thank Pierre Delplace for discussion on the discriminant in the previous work~\cite{Delplace_Resul_arXiv21}.
This work is supported by JSPS Grant-in-Aid for Scientific Research on Innovative Areas ``Discrete Geometric Analysis for Materials Design": Grants No.~JP17H06469 and No.~JP20H04627.
This work is also supported by JSPS KAKENHI Grants No.~JP17H06138, No.~JP20K14371, and No.~JP21K13850.

%


\begin{thebibliography}{74}
\bibitem [{EP3()}]{EP3Dind_ftnt}%
  \BibitemOpen
  \href@noop {} {}\bibinfo {note} {Suppose that an exceptional loop extends
  from $k_z=-\pi$ to $\pi$. In such a case, computing $\zeta_{\mathrm{I},2D}$
  for $k_z=0,\pi$ captures this loop while $\zeta_{\mathrm{I},3D}$ takes
  $1$}\BibitemShut {NoStop}%
\end{thebibliography}

\appendix
\section{
Details of the discriminant
}
\label{sec: Disc app}

Consider a polynomial
\begin{eqnarray}
f(x)&=& a_Nx^N+a_{N-1}x^{N-1}+\ldots+a_1x+a_0,
\end{eqnarray}
with $a_l\in \mathbb{C}$.
The discriminant of $f(x)$ ($\mathrm{Disc}[f(x)]$) is proportional to the resultant of $f(x)$ and $f'(x):=\partial_x f(x)$
\begin{eqnarray}
(-1)^{N(N-1)/2}a_N \mathrm{Disc}[f(x)] &=& \mathrm{Res}[f(x), f'(x)].
\end{eqnarray}
Here, $\partial_x$ denotes derivative respect to $x$.
Because the resultant $\mathrm{Res}[f(x), f'(x)]$ is defined as the determinant of the Sylvester matrix, 
\begin{eqnarray}
&&\mathrm{Res}[f(x), f'(x)] \nonumber \\
&&=\mathrm{det}
\left(
\begin{array}{cccccccc}
a_0    & 0       & \cdots & 0       & b_1      & 0             & \cdots & 0 \\
a_1    & a_0     & \ddots & \vdots  & b_2      & b_1           & \ddots & \vdots \\
a_2    & a_1     & \ddots & a_0     & b_3      & b_2           & \ddots & 0 \\
\vdots & \vdots  & \ddots & a_1     & \vdots   & \vdots        & \ddots & 0 \\
a_N    & a_{N-1} & \cdots & a_2     & b_N      & b_{N-1}       & \cdots & b_1 \\
0      & a_N     & \ddots & \vdots  & 0        & b_N           & \ddots & b_2 \\
\vdots & \vdots  & \ddots & a_{N-1} & \vdots   & \vdots        & \ddots & \vdots \\
0      & 0       & \cdots & a_N     & 0        & 0             & \cdots & b_N
\end{array}
\right),\nonumber \\
\end{eqnarray}
with $b_l=la_l$, the discriminant can be written in terms of $a_l$ ($l=0,\ldots,N$).

Specifically, for $N=3$, we have
\begin{eqnarray}
&&\mathrm{Res}[f(x), f'(x)] \nonumber \\
&&=\mathrm{det}
\left(
\begin{array}{ccccc}
a_0 & 0   & b_1 & 0   & 0   \\
a_1 & a_0 & b_2 & b_1 & 0   \\
a_2 & a_1 & b_3 & b_2 & b_1 \\
a_3 & a_2 &  0  & b_3 & b_2 \\
0   & a_3 &  0  & 0   & b_3
\end{array}
\right).
\end{eqnarray}
Thus, the discriminant is written as
\begin{eqnarray}
\mathrm{Disc}[f(x)]&=& - 4a_1^3 a_3 - 27 a_0^2 a_3^2 +a_1^2 a_2^2 \nonumber \\
&& \quad \quad + 18 a_0 a_1 a_2 a_3 -4 a_0 a_2^3.
\end{eqnarray}

\section{
Relation between symmetry indicators
}
\label{sec: symm ind app}
Parity eigenvalues are implicitly involved with indicators defined in Eqs.~(\ref{eq: nu const inv 2D})~and~(\ref{eq: nu const inv 3D}). 
In addition, exceptional points captured by these indicators correspond to the gapless nodes in Hermitian systems with chiral symmetry.

Firstly, let us consider the following Hamiltonian 
\begin{eqnarray}
\label{eq: h of delta}
\tilde{h}(\bm{k})&=& 
\left(
\begin{array}{cc}
0 & \Delta(\bm{k}) \\
\Delta^*(\bm{k}) & 0
\end{array}
\right).
\end{eqnarray}
The exceptional point emerge when $\Delta(\bm{k})=0$ holds which is described by zero modes of Eq.~(\ref{eq: h of delta}).
We note that $\tilde{h}(\bm{k})$ satisfies
\begin{subequations}
\begin{eqnarray}
\label{eq: h of delta inv}
\sigma_1 \tilde{h}(\bm{k}) \sigma_1&=& \tilde{h}(-\bm{k}), \\
\label{eq: h of delta chiral}
\sigma_3 \tilde{h}(\bm{k}) \sigma_3&=& -\tilde{h}(\bm{k}), 
\end{eqnarray}
\end{subequations}
which means that $\tilde{h}(\bm{k})$ preserves inversion and chiral symmetry.
Equation~(\ref{eq: h of delta inv}) holds due to Eq.~(\ref{eq: Delta inv dag}).

We note that $\Delta(\bm{k})$ is real at TRIM. 
In addition, Eq.~(\ref{eq: h of delta inv}) indicates that the sign of $\Delta(\bm{k})$ corresponds to the parity eigenvalue for the occupied state (i.e., the eigenstate of $\tilde{h}(\bm{k})$ whose eigenvalue is negative). 
In this sense, Eqs.~(\ref{eq: nu const inv 2D})~and~(\ref{eq: nu const inv 3D}) implicitly compute parity eigenvalues of occupied states.

Correspondence between the exceptional points and gapless excitations for Hermitian systems can be seen in the following relation
\begin{subequations}
\label{eq: zeta and z app}
\begin{eqnarray}
\zeta_{\mathrm{I},2D}&=& (-1)^{z_{\mathrm{I},4}},\\
\zeta_{\mathrm{I},3D}&=& (-1)^{z_{\mathrm{I},8}},
\end{eqnarray}
\end{subequations}
where indicators $z_{\mathrm{I}4}$ and $z_{\mathrm{I}8}$ are defined below.
Noting that the indicators $z_{\mathrm{I}4(8)}$ predicts gapless excitations in the two- (three-) dimensional Hamiltonian~(\ref{eq: h of delta})~\cite{Ono_SymmInd-Hermichiral_PRB18,Shiozaki_SymmInd_PRB21}, we can see that exceptional points correspond to these gapless excitations.

Indicators $z_{\mathrm{I}4}$ and $z_{\mathrm{I}8}$ are defined as follows.
Consider a Hermitian system satisfying 
\begin{eqnarray}
U_{\mathrm{I}} \tilde{h}_0(\bm{k}) U_{\mathrm{I}} &=& \tilde{h}_0(-\bm{k}), \\
U_{\mathrm{C}} \tilde{h}_0(\bm{k}) U_{\mathrm{C}} &=& -\tilde{h}_0(\bm{k}),
\end{eqnarray}
with unitary matrices which satisfy $U^2_{\mathrm{I}}=U^2_{\mathrm{C}}=\1$ and $U_{\mathrm{I}}U_{\mathrm{C}}=-U_{\mathrm{C}}U_{\mathrm{I}}$.
Symmetry indicator of such a system is discussed in  Refs.~\onlinecite{Ono_SymmInd-Hermichiral_PRB18,Shiozaki_SymmInd_PRB21}.
In two dimensions, the $\mathbb{Z}_4$-indicator is defined as
\begin{eqnarray}
z_{\mathrm{I}4}&:=& \frac{1}{2}\sum_{k\in\mathrm{TRIM}} (N^{+}_{\bm{k}}-N^{-}_{\bm{k}})  \quad (\mathrm{mod}4),
\end{eqnarray}
where $N^{+(-)}_{\bm{k}}$ denotes the number of occupied bands with the positive (negative) parity at $\bm{k}\in \mathrm{TRIM}$.
For $z_{\mathrm{I}4}=1,3$, the system is gapless due to the chiral symmetry~\cite{Ono_SymmInd-Hermichiral_PRB18,Shiozaki_SymmInd_PRB21}. 
In three dimensions, the $\mathbb{Z}_8$-indicator is defined as
\begin{eqnarray}
z_{\mathrm{I}8}&:=& \frac{1}{2}\sum_{k\in \mathrm{TRIM}} (N^{+}_{\bm{k}}-N^{-}_{\bm{k}}) \quad  (\mathrm{mod}8).
\end{eqnarray}
For odd $z_{\mathrm{I}8}$, the system shows gapless lines due to the chiral symmetry.

Now, we prove Eq.~(\ref{eq: zeta and z app}) by focusing on the two-dimensional case.
Firstly, we note that $N^+_{\bm{k}}+N^-_{\bm{k}}=N_{\mathrm{occ.}}$ holds where $N_{\mathrm{occ.}}$ denotes the number of occupied states.
Thus, we have $z_{\mathrm{I},4}=2N_{\mathrm{occ.}}-\sum_{\bm{\Gamma}_j} N^{-}_{\bm{\Gamma}_j}$, which results in 
\begin{eqnarray}
(-1)^{z_{\mathrm{I},4}} &=& \prod_{\bm{\Gamma}_j\in \mathrm{TRIM}} (-1)^{N^{-}_{\bm{\Gamma}_j}}.
\end{eqnarray}
Applying the above relation to the Hamiltonian~(\ref{eq: h of delta}), we have
\begin{eqnarray}
(-1)^{z_{\mathrm{I},4}} &=& \prod_{\bm{\Gamma}_j\in \mathrm{TRIM}} \mathrm{sgn}\Delta(\bm{\Gamma}_j),
\end{eqnarray}
which is equivalent to Eq.~(\ref{eq: zeta and z app}a).
In a similar way we can prove Eq.~(\ref{eq: zeta and z app}b).

The above results elucidate that exceptional points captured by the discriminant indicators correspond to gapless excitations of $\tilde{h}(\bm{k})$ captured by indicators for Hermitian systems.

\section{
Eq.~(\ref{eq: THT=H^* and IHI=H}) $\Rightarrow $ Eq.~(\ref{eq: const. on Delta SPER})
}
\label{sec: symm const app}
Supposing that Eq.~(\ref{eq: THT=H^* and IHI=H}) holds, we can prove Eq.~(\ref{eq: const. on Delta SPER}) as follows.
Firstly, we note that Eq.~(\ref{eq: THT=H^* and IHI=H}) connects the energy eigenvalues at $\bm{k}$ and at $-\bm{k}$
\begin{subequations}
\label{eq: e(k)=e(-k) ordinary TRS and inv}
\begin{eqnarray}
 \epsilon_n(-\bm{k}) &=& \epsilon^*_n(\bm{k}), \\
 \epsilon_n(\bm{k}) &=& \epsilon_n(-\bm{k}).
\end{eqnarray}
\end{subequations}
As the discriminant is computed from the eigenvalues [see Eq.~(\ref{eq: Defs of disc})], we obtain Eq.~(\ref{eq: const. on Delta SPER}).

In the following, we prove Eq.~(\ref{eq: e(k)=e(-k) ordinary TRS and inv}).
The condition of the time-reversal symmetry results in Eq.~(\ref{eq: e(k)=e(-k) ordinary TRS and inv}a), which can be seen as follows.
Suppose that $|R_n(\bm{k})\rangle$ are right eigenvectors with eigenvalue $\epsilon_n(\bm{k})$. Then, we have
\begin{eqnarray}
H(-\bm{k}) U_{\mathrm{T}} \mathcal{K}  |R_n(\bm{k})\rangle&=& U_{\mathrm{T}} H^*(\bm{k}) \mathcal{K} |R_n(\bm{k})\rangle \nonumber \\
                                              &=& U_{\mathrm{T}} \mathcal{K} H(\bm{k}) |R_n(\bm{k})\rangle \nonumber \\
                                              &=& \epsilon^{*}_n(\bm{k})U_{\mathrm{T}} \mathcal{K} |R_n(\bm{k})\rangle.
\end{eqnarray}
Here, in the first line we have used Eq.~(\ref{eq: THT=H^* and IHI=H}a). 
Thus, we obtain Eq.~(\ref{eq: e(k)=e(-k) ordinary TRS and inv}a).

In a similar way we can prove Eq.~(\ref{eq: e(k)=e(-k) ordinary TRS and inv}b). Because the relation 
\begin{eqnarray}
H(-\bm{k})U_{\mathrm{I}} |R_n(\bm{k})\rangle &=& U_{\mathrm{I}} H(\bm{k}) |R_n(\bm{k})\rangle \nonumber \\
                                  &=& \epsilon_n(\bm{k})  U_{\mathrm{I}} |R_n(\bm{k})\rangle,
\end{eqnarray}
holds, we obtain Eq.~(\ref{eq: e(k)=e(-k) ordinary TRS and inv}b).
Here, in the first line, we have used Eq.~(\ref{eq: THT=H^* and IHI=H}b). 

The above facts prove that Eq.~(\ref{eq: const. on Delta SPER}) holds when the Hamiltonian satisfies Eq.~(\ref{eq: THT=H^* and IHI=H}).

\end{document}